\documentclass[eqsecnum,showpacs,amsmath,aps,nofootinbib,superscriptaddress]{revtex4}  
\usepackage{epsfig}
\usepackage{amssymb}
\usepackage[dvips]{color}
\usepackage[latin1]{inputenc}
\usepackage{graphicx}
\usepackage{gensymb}
\usepackage{amsmath}
\usepackage{subfig}
\usepackage{relsize}
 \usepackage{mathtools}
 \usepackage{alphalph}

\def\beq{\begin{equation}}
\def\eeq{\end{equation}}
\def\beqq{\begin{eqnarray}}
\def\eeqq{\end{eqnarray}}

\def\O{{\rm O}}

\newcommand{\bdm}{\begin{displaymath}}
\newcommand{\edm}{\end{displaymath}}

\def\pmb#1{\setbox0=\hbox{$#1$}%
  \kern-.025em\copy0\kern-\wd0
  \kern.05em\copy0\kern-\wd0
  \kern-.025em\raise.0433em\box0}

\begin{document}
\title{Quantum time delay in the gravitational field of a rotating mass} 

\author{Emmanuele Battista}
\email[E-mail: ]{ebattista@na.infn.it}
\affiliation{Istituto Nazionale di Fisica Nucleare, Sezione di Napoli, Complesso Universitario di Monte
S. Angelo, Via Cintia Edificio 6, 80126 Napoli, Italy}

\author{Angelo Tartaglia}
\email[E-mail: ]{angelo.tartaglia@polito.it}
\affiliation{Politecnico di Torino, Corso Duca degli Abruzzi 24, 10129 Torino, Italy and ISBM}

\author{Giampiero Esposito}
\email[E-mail: ]{gesposit@na.infn.it}
\affiliation{Istituto Nazionale di Fisica Nucleare, Sezione di
Napoli, Complesso Universitario di Monte S. Angelo, 
Via Cintia Edificio 6, 80126 Napoli, Italy}

\author{David Lucchesi}
\email[E-mail: ]{david.lucchesi@iaps.inaf.it}
\affiliation{INAF/IASP and
INFN, Sezione di Torino, Via Pietro Giuria 1, 10125 Torino, Italy}

\author{Matteo Luca Ruggiero}
\email[E-mail: ]{matteo.ruggiero@polito.it}
\affiliation{Politecnico di Torino, Corso Duca degli Abruzzi 24, 10129 Torino, Italy}

\author{Pavol Valko}
\email[E-mail: ]{pavol.valko@stuba.sk}
\affiliation{Slovak University of Technology in Bratislava}

\author{Simone Dell'Agnello}
\email[E-mail: ]{simone.dellagnello@lnf.infn.it}
\affiliation{Istituto Nazionale di Fisica Nucleare, Laboratori Nazionali di Frascati,
00044 Frascati, Italy}

\author{Luciano Di Fiore}
\email[E-mail: ]{luciano.difiore@na.infn.it}
\affiliation{Istituto Nazionale di Fisica Nucleare, Sezione di Napoli,
Complesso Universitario di Monte S. Angelo,
Via Cintia Edificio 6, 80126 Napoli, Italy}

\author{Jules Simo}
\email[E-mail: ]{JSimo@uclan.ac.uk}
\affiliation{Aerospace Engineering, Computing \& Technology Building, School of Engineering, University of Central Lancashire, Preston, PR1 2HE, United Kingdom}

\author{Aniello Grado}
\email[E-mail: ]{agrado@na.astro.it}
\affiliation{INAF, Osservatorio Astronomico di Capodimonte, 80131 Napoli, Italy}

\date{\today}

\begin{abstract}
We examine quantum corrections of time delay arising in the gravitational field of a spinning oblate source. Low-energy quantum effects occurring in Kerr geometry are derived within a framework where general relativity is fully seen as an effective field theory. By employing such a pattern, gravitational radiative modifications of Kerr metric are derived from the energy-momentum tensor of the source, which at lowest order in the fields is modelled as a point mass. Therefore, in order to describe a quantum corrected version of time delay in the case in which the source body has a finite extension, we introduce a hybrid scheme where quantum fluctuations affect only the monopole term occurring in the multipole expansion of the Newtonian potential. The predicted quantum deviation from the corresponding classical value turns out to be too small to be detected in the next future, showing that new models should be examined in order to test low-energy quantum gravity within the solar system. 
\end{abstract}

\pacs{04.60.Ds, 95.10.Ce}
\maketitle

\section{Introduction}

The analysis of timelike and null geodesics evolving in Schwarzschild geometry leads to four important predictions of general relativity within the solar system which represented, in the last century, those crucial challenges that Einstein theory had to overcome before being thoroughly accepted, i.e., the gravitational redshift, the precession of planetary orbits, the bending of light, and the time delay of radar signals. The latter is known also as Shapiro time delay effect (or gravitational time delay effect) and was proposed in 1964 \cite{Shapiro64}. Upon assuming a static and spherically symmetric body, its exterior gravitational field can be described by the Schwarzschild metric \cite{Schwarzschild}, and this makes it possible to measure the increase in time delay between the transmission of an electromagnetic signal towards either of the inner planets and the detection of the echoes. Such a phenomenon is mainly due to the well know relativistic result according to which the speed of light depends on the strength of the gravitational field encountered along its path, whereas the contribution arising from the deflection of its trajectory is negligible, being of order ${\rm O}(G^2 M^2/c^6)$ (see Eq. (\ref{ShapiroTimeDelay_Schwarzschild})). Bearing in mind the above considerations, we can consider a quasi-Cartesian coordinate system at the post-Newtonian level with origin in the central (deflecting and) delaying body and suppose that the propagation occurs in the $z=0$ plane (coincident with the plane of ecliptic) so that the squared line element can be written as
\begin{equation}
{\rm d}s^2 = c^2 {\rm d} \tau^2 = g_{00} c^2 {\rm d}t^2 + g_{xx}{\rm d}x^2+ g_{yy} {\rm d}y^2,
\label{Schwarzschild_line_element}
\end{equation}
where $c$ represents the speed of light, $\tau$ the invariant proper time and\footnote{We are adopting harmonic gauge. }
\begin{equation}
\begin{split}
g_{00} & = - \left(1+2 \mathcal{U} \right) + {\rm O}(c^{-4}), \\
g_{xx} &=g_{yy} = \left( 1- 2 \mathcal{U} \right) +{\rm O}(c^{-4}),
\label{Schwarzschild_metric_components}
\end{split}
\end{equation}
with
\begin{equation}
\mathcal{U}=-\dfrac{GM}{c^2 r},
\end{equation}
$M$ being the mass of the central body, $G$ the Newtonian gravitational constant and $r$ the Euclidean distance in the reference plane. Moreover, by considering the propagation of an electromagnetic wave (i.e., a null geodesic) Eq. (\ref{Schwarzschild_line_element}) can be solved for the time element ${\rm d}t$, so that if we further employ the standard post-Newtonian approximation (see \cite{MTW} and also \cite{Tartaglia2002,Tartaglia2016} and the references therein) where the beam travels along the straight path $x=b={\rm const.}$, the emitter is located at $(x=b;y=-y_1)$, the receiver at $(x=b;y=y_2)$ (with $y_1,y_2>0$, $b\ll y_1,b \ll y_2$, and $y_1 \simeq y_2$), then the time propagation measured by a central-body-based clock reads as
\begin{equation}
\Delta t_{{\rm prop}}= \dfrac{y_2+y_1}{c}+\dfrac{4 G M}{c^3} \log \left(\dfrac{y_2+y_1}{b}\right) + {\rm O}(G^2M^2/c^6).
\label{ShapiroTimeDelay_Schwarzschild}
\end{equation}
The first term in Eq. (\ref{ShapiroTimeDelay_Schwarzschild}) accounts for the time propagation in a flat spacetime, the second term represents the Shapiro time delay and the second order terms $ {\rm O}(G^2M^2/c^6)$ indicate corrections involving the deflection of light path. 

A more realistic model can be built up by considering a rotating body whose gravitational field is described in terms of Kerr metric \cite{Kerr}. In this case further contributions to (\ref{ShapiroTimeDelay_Schwarzschild}) emerge due to spin and quadrupole moment of the body (see Eq. (\ref{ShapiroTimeDelay_Kerr})). Moreover, for this geometry Einstein theory predicts the occurrence of other peculiar phenomena involving test particles, gyroscopes, photons and clocks \cite{Ciufolini-Wheeler}. Indeed, a particle co-rotating with the spinning object has an orbital period longer than the one of a particle counter-rotating at the same distance from the central body. Furthermore, the orbital plane of a mass orbiting around the twisting source is dragged around in the sense singled out by its rotation. Therefore, small gyroscopes determining the orientation of axes of a local freely falling frame are forced to rotate with respect to the direction identified by distant stars. Such a phenomenon was named by Einstein himself as ``(inertial) frame dragging'' and it is also known as Lense-Thirring effect \cite{LT}. When we deal with situations affecting photons and clocks, another interesting effect is represented by the fact that a photon co-rotating with the central mass takes less time to return to a fixed point than a photon counter-rotating: all around a closed path encompassing the spinning body the synchronization of clocks is unachievable. 

\begin{figure}
\includegraphics[scale=0.45]{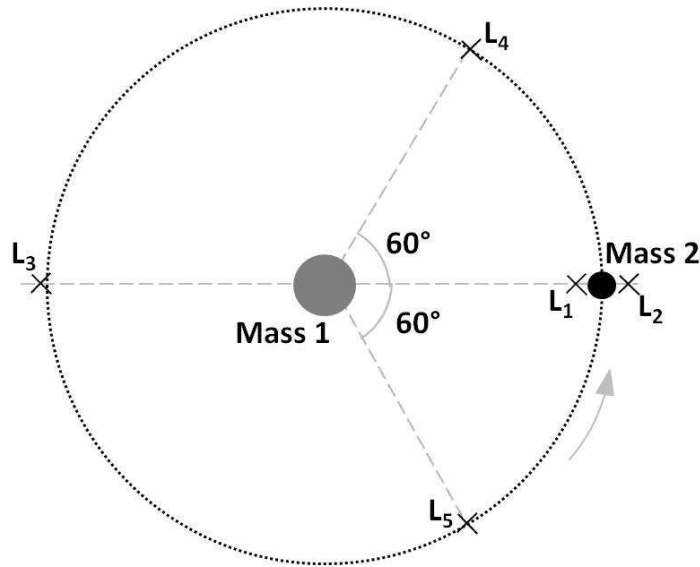}
\caption{Schematic view of Lagrangian points in the restricted three-body system. The masses depicted are refereed to as primaries (in this paper the Sun and the Earth); the points $L_1$, $L_2$, and $L_3$ are also called collinear libration points, while $L_4$ and $L_5$, which according to Newtonian theory are expected to be located at the same distance from the primaries, are known as triangular or non-collinear points.}
\label{LagP.eps}
\end{figure}
Although Shapiro time delay could sound as an aged effect, it can still represent a crucial relativistic phenomenon for fundamental physics experiments at the scale of the inner solar system thanks to the ideas developed in Refs. \cite{Tartaglia2002,Tartaglia2016}, where an interesting proposal, nicknamed ``LAGRANGE'', has been set forth. By exploiting the fact that Lagrangian points of the Sun-Earth system (see Fig. \ref{LagP.eps}) form a configuration rigidly rotating with the primaries \cite{Pars,B1,BattistaThesis}, it is possible to conceive an experimental pattern where four spacecraft, located in $L_1$, $L_2$, $L_4$, and $L_5$, reciprocally exchange  electromagnetic signals. The aim of ``LAGRANGE'' consists in exploiting the time of flight measurements of beams travelling along various paths having Lagrangian points as their vertices in order to gain valuable informations about the gravitational and the gravitomagnetic field of the bodies labelled as ``Mass 1'' and  ``Mass 2'' in Fig. \ref{LagP.eps}. In particular, by employing an open path approaching either of the bodies it is possible to measure their quadrupole moment; on the other hand, if a closed contour is adopted the Lense-Thirring effect, considered as a manifestation of gravitomagnetism, can be tested as well \cite{Tartaglia2002,Tartaglia2016}. Moreover, such patterns represent a significant opportunity to constrain the post-Newtonian parameter $\gamma$, espressing the curvature per unit of mass. 

Beside the renewed interest in the classical aspects of general relativity mentioned above, in recent literature a valuable theoretical framework has been proposed in order to deal also with the quantum nature of Einstein theory, at least at low-energy scales like those emerging in our solar system \cite{B1,B1a,BattistaThesis,Donoghue,B2,B3}. The key feature underlying this pattern consists in considering general relativity as an effective field theory. The origin of this model arises from the fact that even in a strictly nonrenormalizable field theory quantum predictions can still be worked out, provided we adopt an effective theory approach. A valuable example of this situation is surely represented by chiral perturbation theory in the context of quantum chromodynamics \cite{Scherer}. Therefore, by adopting this procedure together with the background field quantization it is possible to pursue a consistent quantization program of general relativity, since the troublesome ultraviolet singularities occurring in the traditional renormalization scheme can be absorbed into the phenomenological constants characterizing the effective action of the theory (see Eq. (\ref{eff_Lagrangian})). This amounts to introduce a never ending set of additional higher derivative couplings into the model in such a way that Einstein theory represents now only a minimal theory, i.e., the one marking out the low-energy quantum effects. In fact, in order to treat general relativity as an effective field theory the Einstein-Hilbert Lagrangian (here considered without cosmological term)
\begin{equation}
\mathcal{L}_{\rm GR}=\sqrt{-g} \dfrac{c^4}{16 \pi G}\,  R,
\end{equation}
where $g=\det(g_{\mu \nu})$ and $R=g^{\mu \nu}R_{\mu \nu}$ is the Ricci scalar, should be equipped with all possible higher derivative terms which turn out to be, at the same time, compatible with general covariance requirement and suitable for absorbing any field singularities generated by loop diagrams. Recalling the fact that one-loop singularities occurring in vacuum Einstein theory involve the square of both Ricci scalar and Ricci tensor \cite{'tHooft1974}, this means that the effective action for pure general relativity should read as
\begin{equation}
\mathcal{L}_{\rm eff}=\sqrt{-g} \left( \dfrac{c^4}{16 \pi G} \,R +c_1 R^2 + c_2R_{\mu \nu} R^{\mu \nu} + \dots \right),
\label{eff_Lagrangian}
\end{equation}
 where $c_1$ and $c_2$ are seen as experimentally determined quantities and ellipsis denote higher order couplings, indicating that Eq. (\ref{eff_Lagrangian}) represents an infinite energy power series. As pointed out before, the low-energy regime of quantum gravity is completely ruled by the Einstein-Hilbert sector of (\ref{eff_Lagrangian}). In fact, bearing in mind that the curvature involves second order derivatives of the metric, the momentum space correspondence $i \partial_\mu \sim p_\mu$ makes $R$ and $R_{\mu \nu}$ of order $p^2$, while second order terms in the Lagrangian (\ref{eff_Lagrangian}) involving two powers of the curvature are of order $p^4$. Therefore, at long distances, quantities of order $p^4$ can be extremely smaller than those of order $p^2$ or, in other words, all contributions to (\ref{eff_Lagrangian}) beyond the Einstein-Hilbert part have little effect at low-energy scales. In this regime the massless degrees of freedom, being long ranged,  dominate over massive ones. Since the former are characterized, in Fourier space, by a propagator going like $1/q^2$ and hence not Taylor-expandible around $q^2 \simeq 0$, whereas the latter by a propagator having the form $1/(q^2-m^2)$ which clearly results in an analytical quantity around $q^2\simeq 0$, then it easily turns out that analytic contributions to loop diagrams originate from propagation of massive particles in Feynman diagrams, while nonanalytic ones from massless modes. Therefore, within such an approach the calculations are performed by considering only the nonanalytic contributions to Feynman diagrams. In other words, we have integrated the massive degrees of freedom out of the theory by performing the path integral with respect to these fields. At tree-level, this amounts to replace the massive fields by their classical equations of motion and to introduce for the propagator the expansion we mentioned before. Such an expansion results in a power series 
 \begin{equation}
 \dfrac{1}{q^2-m^2}=-\dfrac{1}{m^2}- \dfrac{q^2}{m^4}+\O(q^4/m^6),
 \end{equation}
which clearly breaks down at energy scales $q^2 \simeq m^2$, making the effective theory be naturally equipped with an ultraviolet cut-off scale. 

By employing this framework, it is possible to evaluate quantum corrections to Newtonian potential \cite{D94,D2003} and all their implications in the analysis of the three-body problem \cite{B1,B2,B3,BattistaThesis} or quantum corrections to exact solutions of Einstein field equations, such as Schwarzschild and Kerr metric \cite{Bohr2003} or Reissner-Nordstr\"{o}m and Kerr-Newman ones \cite{D2002}. In particular, some of us  have developed in Refs. \cite{B1,B2,B3,BattistaThesis} a model concerning the quantum description of the position of Lagrangian points in the restricted three-body problem having the Earth and the Moon as primaries and have shown how the deviations from Newtonian regime involve quantities of the order of few millimetres. Driven by such a result, it would be interesting to evaluate a quantum corrected version of Shapiro time delay by exploiting quantum corrections to Kerr metric reported in Ref. \cite{Bohr2003}. This calculation represents the main purpose of this article.

Before proceeding with the main topics of the paper we believe that, although the abovementioned arguments regarding the momentum-space correspondence of the terms occurring in Eq. (\ref{eff_Lagrangian}) are pragmatic (momentum-space methods fulfil a crucial role in the analysis of ultraviolet divergences in quantum field theory in Minkowski space), they deserve a short digression. Indeed, it is widely known that when dealing with a general curved spacetime the homogeneity required for the existence of a {\it global} momentum-space representation is missing. Therefore, we should abandon such useful framework unless the considered geometry is sufficiently homogeneous or it can be treated as a small perturbation of a homogeneous spacetime. However, the joint effect of the smallness of length scales involved in the analysis of ultraviolet divergences of an interacting model and the equivalence principle underlying Einstein theory (emphasizing the possibility of eliminating locally every {\it observable} effect of the gravitational field, i.e., curved spacetimes are approximately flat in sufficiently small regions) is such that Minkowski space techniques could be applied locally in any curved spacetime. Driven by this clever idea, in Ref. \cite{Bunch} a {\it local} momentum-space representation has been introduced near any given point of a general curved spacetime by exploiting the features of Gaussian normal coordinates, which are defined as follows \cite{MTW}. Consider two events $O$ (regarded as the origin) and $P$ of a generic spacetime linked by a geodesic. The event $P$ is specified by the normal coordinates $y^{\mu}=s \xi^{\mu}$, $\xi^{\mu}$ being the unit tangent vector at the origin to the geodesic connecting $O$ and $P$ and $s$ its arc length. Moreover, at the origin the metric tensor reduces to Minkowski metric $\eta_{\alpha \beta}$ and Christoffel symbols vanish. This set of coordinates is valid in the normal neighborhood $N_{O}$ of $O$ in which the geodesics from the origin do not intersect and hence are well suited to the description of physical phenomena occurring at short wavelengths, where ultraviolet divergences emerge. In particular, by introducing a set of normal coordinates having origin at a generic point $x^{\prime}$, then for any point $x$ lying in the normal neighborhood $N_{x^{\prime}}$ of $x^{\prime}$ the Feynman Green function $G(x,x^{\prime})$ of a scalar field of mass $m$ propagating in an aribtrary curved geometry can be written by the means of Fourier transformation as \cite{Bunch}
\begin{equation}
\bar{G}(x,x^{\prime}) \equiv g^{1/4}(x) G(x,x^{\prime})=\int \dfrac{{\rm d}^4\,k}{(2 \pi)^4} {\rm e}^{{\rm i}ky}\, \bar{G}(k), \;\;\;\;\;\; \forall\, x \in N_{x^{\prime}},
\end{equation}
where $ky \equiv \eta^{\alpha \beta} k_{\alpha}y_{\beta}$ and $\bar{G}(k) \equiv \bar{G}(k,x^{\prime})$. The Fourier transform is then showed to be expressible through the asymptotic expansion
\begin{equation}
\bar{G}(k) \sim \left(k^2+m^2\right)^{-1}+f_1(R) \left(k^2+m^2\right)^{-2}+ \dots,
\end{equation}
$f_1(R)$ being a function of the curvature. Furthermore, this approach turns out to be equivalent to the Schwinger-DeWitt proper time representation, according to which the propagator $G(x,x^{\prime})$ reads as \cite{DW}
\begin{equation}
G(x,x^{\prime})= \dfrac{{\rm i}\Delta^{1/2}(x,x^{\prime})}{(4 \pi)^2} \int_{0}^{\infty} \dfrac{{\rm i}{\rm d}s}{({\rm i}s)^2} \exp \left(-{\rm i}m^2 s -\dfrac{\sigma}{2 {\rm i}s} \right) F(x,x^{\prime};{\rm i}s),
\end{equation}
where $\Delta(x,x^{\prime})$ denotes the van Vleck determinant
\begin{equation}
\Delta(x,x^{\prime})=-g^{-1/2}(x) \det [-\partial_{\mu} \, \partial_{\nu^{\prime}} \,\sigma(x,x^{\prime}) ] 	\,g^{-1/2}(x^{\prime}),
\end{equation}
$\sigma(x,x^{\prime})$ represents half the square of the geodesic distance between $x$ and $x^{\prime}$ and $F(x,x^{\prime};{\rm i}s)$ is an amplitude function depending on curvature. Therefore, the above arguments demonstrate how the correspondence $i \partial_\mu \sim p_\mu$ can be still valid {\it locally} in curved spaces, provided we adopt the appropriate pattern. However, since the model developed by the authors of Ref. \cite{Donoghue} makes use of a quantization programme developed in Minkowski background, the above considerations involving energy scales underlying the effective Lagrangian (\ref{eff_Lagrangian}) can be considered to be still well-founded. 

The paper is organised as follows. Section II deals with classical time delay experienced by an electromagnetic signal in the gravitational field of an oblate rotating source of mass $M$ having angular momentum $\boldsymbol{J}$ and quadrupole moment $J_2$. By employing a quasi-Cartesian coordinate system within the standard isotropic post-Newtonian approximation the expression of the propagation time $\Delta t_{\rm prop}$ taken by the beam to follow a straight path is achieved. In Sec. III we perform the calculations underlying the quantum corrected version of the time delay. Such a computation has led us to propose a hybrid scheme where quantum modifications originally conceived for a point mass are applied to an extended source as well. Concluding remarks and open problems are reported in Sec. IV. 

\section{Classical time delay } \label{Sec_Classical_time_delay}

In this section we analize the more realistic model where the delaying body having mass $M$ is not represented simply by a static sphere, but instead by a spinning object whose shape deviates from the spherical one. Such a configuration can be caught by adopting the axially symmetric and stationary Kerr metric \cite{Kerr}, instead of the static (i.e., $g_{0i}=0,\dfrac{\partial}{\partial t}g_{\mu \nu}=0$) spherically symmetric Schwarzschild geometry (\ref{Schwarzschild_line_element}) and (\ref{Schwarzschild_metric_components}). Written in Boyer-Lindquist coordinates $(r,\theta,\phi)$ \cite{Boyer-Lindquist}, Kerr metric reads as
\begin{equation}
{\rm d}s^2= \left(1-\dfrac{r_s\,r}{\delta^2}\right)c^2 {\rm d}t^2 - \dfrac{\delta^2}{\Delta^2} {\rm d}r^2 - \delta^2 {\rm d}\theta^2 -\left(r^2 + \alpha^2 + \dfrac{r_s\,r\,\alpha^2}{\delta^2} \sin^2 \theta \right) \sin^2 \theta \, {\rm d}\phi^2 + 2\dfrac{ r_s\,r\,\alpha}{\delta^2} \sin^2 \theta \; c \,{\rm d}\phi \,{\rm d}t,
\label{Kerr_Metric_Boyer}
\end{equation}
where
\begin{equation}
\begin{split}
\Delta&= r^2-r_s\,r+\alpha^2,\\
\delta^2 &= r^2 + \alpha^2 \cos^2 \theta,
\end{split}
\end{equation}
$r_s$ being the Schwarzschild radius and $\alpha$ a parameter related to the angular momentum $J$ of the body, i.e., 
\begin{equation}
\begin{split}
r_s&=\dfrac{2GM}{c^2}, \\
\alpha & =\dfrac{J}{Mc},
\end{split}
\end{equation}
while the relation between Boyer-Lindquist coordinates and Cartesian ones $(x,y,z)$ is given by
\begin{equation}
\begin{split}
x&=\sqrt{r^2 + \alpha^2} \, \sin \theta \, \cos \phi, \\
y&=\sqrt{r^2 + \alpha^2} \, \sin \theta \, \sin \phi, \\
z&= r \cos \theta.
\end{split}
\end{equation}
The metric (\ref{Kerr_Metric_Boyer}) has two coordinate singularities represented by the conditions $g_{00}=0$ and $g_{11}=\infty$. Unlike Schwarzschild metric, where these two limiting values are achieved simultaneously on the surface $r=r_s$, for Kerr geometry there exist two different surfaces: $g_{00}$ vanishes when $\delta^2=r\,r_s$, while $g_{11}$ diverges when $\Delta=0$. The former circumstance implies a quadratic equation whose larger root is given by
\begin{equation}
r_0=\dfrac{r_s}{2}+\sqrt{\left(\dfrac{r_s}{2}\right)^2-\alpha^2 \cos^2 \theta},
\end{equation}
whereas the latter leads to another quadratic equation having as its larger solution
\begin{equation}
r_{{\rm hor}}=\dfrac{r_s}{2}+\sqrt{\left(\dfrac{r_s}{2}\right)^2-\alpha^2 }.
\end{equation}
It is clear from the above equations that the surface $r=r_{{\rm hor}}$ is a sphere whereas $r=r_0$ represents an oblate figure of rotation containing this sphere and touching it at the poles ($\theta=0,\pi$). The sphere $r=r_{{\rm hor}}$ represents a null hypersurface which does not extend to spatial infinity so that its sections $t=$ constant are closed spatial surfaces. In other words, $r=r_{{\rm hor}}$ represents an event horizon, analogous to Schwarzschild horizon in the spherically symmetric case, which is passed by causal geodesics in only one direction, i.e., toward the interior. The space between $r=r_0$ and the horizon is known as ergosphere and it possesses the nice feature according to which all particles and light rays inside it must be inevitably subjected to rotational motion. Moreover, all particles can emerge from the ergosphere and reach the external region and viceversa \cite{Boyer-Lindquist}. 

If we consider, like in Sec. I, a quasi-Cartesian coordinate system with the central body located at the origin and having angular momentum $\boldsymbol{J}$ along the $z$-axis, representing also its symmetry axis, then in the hypothesis in which the motion takes place in the $z=0$ plane and by employing a standard isotropic post-Newtonian approximation, the line element in Kerr geometry reads as \cite{Tartaglia2016}
\begin{equation}
{\rm d}s^2= c^2 {\rm d}\tau^2 \simeq  g_{00} \,c^2 {\rm d}t^2 + g_{xx}\, {\rm d}x^2+g_{yy}\,{\rm d}y^2 + 2 g_{0x} \,{\rm d}x\,{\rm d}t+ 2 g_{0y} \,{\rm d}y\,{\rm d}t,
\label{Kerr_Metric1}
\end{equation}
where up to first order in angular momentum and in harmonic gauge
\begin{equation}
\begin{split}
g_{00}&=-\dfrac{1+U}{1-U}=-1-2U+\O(c^{-4}), \\
g_{ij}&=\delta_{ij}(1-U)^2 + U^2 \left(\dfrac{1+U}{1-U} \right) \dfrac{r_i r_j}{r^2}=\delta_{ij}\left(1-2U\right)+\O(c^{-4}),\\
g_{0x} & = 2\dfrac{GJ}{c^2 r^3}(-y)+\O(c^{-4}),\\
g_{0y}&= 2\dfrac{GJ}{c^2 r^3}(x)+\O(c^{-4}),\\
\end{split}
\label{Kerr_Metric2}
\end{equation}
$r$ being the Euclidean distance in the reference plane and $U$ the gravitational potential
\begin{equation}
U \simeq -\dfrac{GM}{c^2 r}\left[1+\dfrac{J_2}{2}\left(\dfrac{R}{r}\right)^2 \right] \equiv-\dfrac{GM}{c^2 r}\left(1+ \chi(r) \right)
\label{potential_U}
\end{equation}
 where $R$ represents the (equatorial) radius of the body and $J_2$ its quadrupole moment taking into account its oblateness. Moreover the non-diagonal part $g_{0i}$ of the metric tensor (i.e., the gravitomagnetic potential) is a solution of the Poisson equation \cite{Ciufolini-Wheeler}
 \begin{equation}
 \triangle V_i = -4 \pi G \rho v_i,
 \end{equation}
 where $g_{0i} \equiv -4 V_i/c^3$, $\rho$ represents the mass density of the stationary source whereas $\rho {\boldsymbol v}$ its mass-current density. 
 
 If we examine the propagation of an electromagnetic beam along the straight line $x=b=$ constant, then Eqs. (\ref{Kerr_Metric1}) and (\ref{Kerr_Metric2}) reduce to 
 \begin{equation}
0 \simeq g_{00} \,c^2 {\rm d}t^2 +g_{yy}\,{\rm d}y^2 +  2 g_{0y} \,{\rm d}y\,{\rm d}t,
\end{equation}
which, solved for ${\rm d}t$, gives
\begin{equation}
{\rm d}t = - \dfrac{g_{0y}+\sqrt{g_{0y}^2-c^2g_{00}g_{yy}}}{c^2g_{00}}\, {\rm d}y.
\label{classical_dt}
\end{equation} 
 By introducing the expansions
 \begin{equation}
 \begin{split}
 & \dfrac{1}{g_{00}}=-1+2U+\O(c^{-4}),\\
 & \sqrt{g_{0y}^2-c^2g_{00}g_{yy}} =c+\O(c^{-3}),
 \end{split}
 \label{expansion1}
 \end{equation}
 it is easy to write Eq. (\ref{classical_dt}) in the form
 \begin{equation}
{\rm d}t= {\rm d}y \left(\dfrac{1}{c}-\dfrac{2U}{c}+\dfrac{g_{0y}}{c^2}+\O(c^{-6}) \right).
 \end{equation}
At this stage, we can consider once again the situation in which the emitter is located at $y=-y_1$, the receiver at $y=y_2$ (with $y_1$ and $y_2$ both positive quantities, $y_1\gg b$, $y_2 \gg b$, $y_1 \simeq y_2$), and hence by integrating the above expression jointly with the results of Appendix \ref{Appendix_Integrals} we easily obtain
\begin{equation}
\Delta t_{{\rm prop}}= \dfrac{y_2+y_1}{c}+\dfrac{4 G M}{c^3} \log \left(\dfrac{y_2+y_1}{b}\right)+\dfrac{2GM}{c^3}\left(\dfrac{R}{b}\right)^2J_2 \pm  \dfrac{4GJ}{c^4b}+ {\rm O}(c^{-5}).
\label{ShapiroTimeDelay_Kerr}
\end{equation}
The first two terms appearing in Eq. (\ref{ShapiroTimeDelay_Kerr}) are the same as those occurring in Eq. (\ref{ShapiroTimeDelay_Schwarzschild}): the new ingredients  are represented by the last two factors. The third contribution in (\ref{ShapiroTimeDelay_Kerr}) is related to the oblateness of the central mass $M$, whereas the last term arises from its angular momentum $\boldsymbol{J}=J \hat{z}$, the $\pm$ sign depending on chirality so that if the beam propagates in the same sense of rotation of the spinning body we will have the positive sign, contrarily the minus sign will come out. However, note that in the weak-field and slow-motion limit (i.e., $\dfrac{J}{Mcr} \ll 1$ and $\dfrac{GM}{c^2 r} \ll1$) the delay effect due to spin (gravitomagnetic delay) decouples from quadrupole moment time delay.

At this stage we can consider a path linking Sun-Earth Lagrangian points $L_1$ and $L_2$. This means that the role of the central delaying and deflecting body will be fulfilled by the Earth. By taking into account this configuration, it will be possible (at least in principle) to obtain a measurement of the Earth's quadrupole coefficient in a way independent of the usual space geodesy techniques based on the inter-satellite tracking and from the precise orbit determination of laser-ranged satellites in orbit at a relatively high altitude \cite{Tartaglia2016, Reigber}. The distance between Sun-Earth $L_1$ and $L_2$ is such that $y_2+y_1 \simeq 3 \times 10^9 \,{\rm m}$, thus by assuming an impact parameter of the order of the Earth's radius, i.e., $b \simeq 6.4 \times 10^6 \,{\rm m}$, from Eq. (\ref{ShapiroTimeDelay_Kerr}) we obtain
\begin{equation}
\Delta t_{{\rm prop}} \simeq (10\,{\rm s})+( 3.6 \times 10^{-10}\,{\rm s})+( 3.2 \times 10^{-14}\, {\rm s})+( \pm 3 \times 10^{-17}\,{\rm s}).
\end{equation}
It is thus clear that the Earth quadrupole moment could be measured within this framework once a round trip travel for propagation time is chosen, since in this configuration the chiral contribution from the gravitomagnetic field cancels out, provided that the path of the beam always lies on the same side of the Earth. In fact, the knowledge of the oblateness of the Earth is particularly important in the analysis of its internal structure and mass distribution and the induced long-term variations. Phenomena like the melting from glaciers and ice sheets as well as mass changes in the oceans and in the atmosphere are responsible for variations in the rate of the global mass redistribution with a consequent time dependency in the quadrupole coefficient characterized by annual and interannual variations. This kind of measurement can be initiated by Earth, the delaying body, with all the advantages of an Earth based Laboratory equipped with the best time-measuring apparatus
to perform the experiment \cite{Tartaglia2016}. 
 
\section{Quantum corrected time delay}

Having seen the classical time delay occurring in the field of a rotating mass, we are now ready to investigate how quantum theory, developed within the context in which general relativity is seen as an effective field theory, affects it.

As we know, the gravitational field surrounding a body is described by the spacetime metric $g_{\mu \nu}$ which solves Einstein equations. Quantum corrections to Kerr metric are developed through a pattern where $g_{\mu \nu}$ is derived from the energy momentum tensor $T_{\mu \nu}$ of the source whose dynamics is ruled by Einstein equations \cite{Bohr2003}. This framework is inspired by the hybrid scheme where quantum matter yields, in the low-energy regime, effects which dominate those arising from quantum gravity. By introducing  the field expansion
\begin{equation}
g_{\mu \nu}= \eta_{\mu \nu} + h^{(1)}_{\mu \nu}+h^{(2)}_{\mu \nu}+\dots,
\end{equation}
the superscript referring to powers of the gravitational coupling $G$, it is possible to show that, to first order, $T_{\mu \nu}$ represents the energy-momentum tensor of a point particle. At the following order in the fields, $T_{\mu \nu}$ denotes the energy and momentum carried by the gravitational field $h^{(1)}_{\mu \nu}$ surrounding the point mass. These corrections to the lowest order result can be read off by employing a one-loop Feynman-diagram analysis. The masslessness of the graviton ensures the presence of nonanalytic/long-ranged quantum effects in the Kerr metric. It is brilliantly shown in Ref. \cite{Bohr2003} that the computation of the underlying one-loop Feynman diagrams leads to the neat result involving metric tensor components (again in the harmonic gauge and up to first order terms in the angular momentum $\boldsymbol{J}$)
\begin{equation}
\begin{split}
& g_{00}= -1 +\dfrac{2GM}{c^2r}-\dfrac{2G^2M^2}{c^4 r^2}-\dfrac{62G^2M \hbar}{15 \pi c^5 r^3}+\O(c^{-6}),\\
& g_{ij}= \delta_{ij} \left(1+\dfrac{2GM}{c^2r}+\dfrac{G^2M^2}{c^4 r^2}+\dfrac{14G^2M \hbar}{15 \pi c^5 r^3}+\O(c^{-6})\right)+\dfrac{r_i r_j}{r^2} \left( \dfrac{G^2M^2}{c^4 r^2}+\dfrac{76G^2M \hbar}{15 \pi c^5 r^3}+\O(c^{-6})\right),\\
& g_{0i}= \left(\dfrac{2G}{c^2 r^3 }-\dfrac{2G^2M}{c^4r^4}+\dfrac{36G^2 \hbar}{15 \pi c^5 r^5} +\O(c^{-6}) \right) \left( \boldsymbol{J} \times \boldsymbol{r} \right)_{i}, 
\label{Bohr_Quantum_Kerr}
\end{split}
\end{equation}
where $\boldsymbol{r}=(x,y,z)$. 

From the above considerations it should be clear that quantum effects occurring in Eq. (\ref{Bohr_Quantum_Kerr}) arise from the corrections affecting the energy-momentum tensor of a point mass. Therefore, in order to deal with an extended source having quadrupole moment $J_2$, we adopt the hybrid scheme according to which the quantum pattern (\ref{Bohr_Quantum_Kerr}) affects only the monopole term $M$ occurring in the potential (\ref{potential_U}). In order to include such terms, we write the classical Kerr metric up to second order terms in $U$. Such terms will be written approximatively by using
\begin{equation}
U^2 \simeq \dfrac{G^2M^2}{c^4r^2}(1+ 2 \chi),
\label{U^2_approx}
\end{equation}
i.e., we are neglecting quadratic terms in the quadrupole moment. Therefore, instead of the expressions (\ref{Kerr_Metric2}), the underlying classical theory will be characterized by
\begin{equation}
\begin{split}
g_{00}&=-\dfrac{1+U}{1-U}=-1 + \dfrac{2GM}{c^2 r} (1+\chi)-\dfrac{2G^2 M^2}{c^4 r^2}(1+2\chi)+\O(c^{-6}), \\
g_{ij}&=\delta_{ij}(1-U)^2 + U^2 \left(\dfrac{1+U}{1-U} \right) \dfrac{r_i r_j}{r^2}=\delta_{ij}\left[1+\dfrac{2GM}{c^2r}(1+\chi)+\dfrac{G^2M^2}{c^4r^2}(1+2\chi)\right]+\dfrac{r_i r_j}{r^2}\dfrac{G^2M^2}{c^4r^2}+\O(c^{-6}),\\
g_{0i} & = \left(\dfrac{2G}{c^2 r^3} -\dfrac{2 G^2 M}{c^4 r^4} + \O(c^{-6})\right)(\boldsymbol{J} \times \boldsymbol{r})_i.
\end{split}
\label{Kerr_Metric3}
\end{equation}
Thus, the joint action of Eqs. (\ref{Bohr_Quantum_Kerr}) and (\ref{Kerr_Metric3}) within the hybrid scheme mentioned before gives rise to the quantum corrected Kerr metric
\begin{equation}
\begin{split}
g_{00}&=-1 + \dfrac{2GM}{c^2 r} (1+\chi)-\dfrac{2G^2 M^2}{c^4 r^2}(1+2\chi)-\dfrac{62 G^2 M \hbar}{15 \pi c^5 r^3} +\O(c^{-6}), \\
g_{ij}&=\delta_{ij}\left[1+\dfrac{2GM}{c^2r}(1+\chi)+\dfrac{G^2M^2}{c^4r^2}(1+2\chi) + \dfrac{14 G^2 M \hbar}{15 \pi c^5 r^3}\right] 
 +\dfrac{r_i r_j}{r^2}\left( \dfrac{G^2M^2}{c^4r^2}+\dfrac{76G^2M\hbar}{15 \pi c^5 r^3}+\O(c^{-6})\right),\\
g_{0i} & = \left(\dfrac{2G}{c^2 r^3} -\dfrac{2 G^2 M}{c^4 r^4} + \dfrac{36G^2\hbar}{15 \pi c^5 r^5}+\O(c^{-6})\right)(\boldsymbol{J} \times \boldsymbol{r})_i.
\end{split}
\label{Quantum_Kerr}
\end{equation}

At this stage, we are ready to perform the calculations leading to quantum time delay by employing the metric tensor components (\ref{Quantum_Kerr}). Like in Sec. \ref{Sec_Classical_time_delay}, we adopt a configuration for which the light ray propagates in the $z=0$ plane following the straight path $x=b$ from the position $y=-y_1$ to $y=y_2$ (with $y_1,y_2 \gg b$, $y_1 \simeq y_2$). Furthermore, the spinning source is located at the origin of the quasi-Cartesian coordinate system chosen and its angular momentum points along the $z$-axis. Thus, the line element describing the null geodesic followed by the photons will be 
\begin{equation}
{\rm d}s^2 =0 \simeq g_{00} \,c^2 {\rm d}t^2 +g_{yy}\,{\rm d}y^2 +  2 g_{0y} \,{\rm d}y\,{\rm d}t,
\end{equation}
which gives for the line element ${\rm d}t$ 
\begin{equation}
{\rm d}t = - \dfrac{g_{0y}+\sqrt{g_{0y}^2-c^2g_{00}g_{yy}}}{c^2g_{00}}\, {\rm d}y.
\label{quantum_dt}
\end{equation} 
In order to evaluate the infinitesimal change in the time coordinate (\ref{quantum_dt}), we need to consider that Eq. (\ref{expansion1}) gets now replaced by the following expansions:
\begin{equation}
\dfrac{1}{c^2 g_{00}}= -\left[\dfrac{1}{c^2}+\dfrac{2GM}{c^4r}(1+\chi)+\dfrac{2G^2M^2}{c^6r^2}(1+2\chi) -\dfrac{62G^2M \hbar}{15 \pi c^7 r^3}+\O(c^{-8})\right],
\end{equation}
\begin{equation}
\sqrt{g_{0y}^2-c^2g_{00}g_{yy}}=c-\dfrac{G^2M^2}{2c^3r^2} (1+2 \chi) + \dfrac{38 G^2 M \hbar }{15 \pi c^5 r^3} + \O(c^{-5}),
\end{equation}
where Eq. (\ref{U^2_approx}) has been exploited and $g_{0y}^2$ has been written as
\begin{equation}
g_{0y}^2= \dfrac{\left(2GJb\right)^2}{c^4r^6}+\O(c^{-6}).
\label{g_0y_squared}
\end{equation} 
Bearing in mind the above relations, Eq. (\ref{quantum_dt}) reads as
\begin{equation}
{\rm d}t= {\rm d}y \left[ \dfrac{1}{c}+\dfrac{2GM}{c^3r}(1+\chi) + \dfrac{2GJb}{c^4r^3}+\dfrac{3G^2M^2}{2c^5r^2}(1+2\chi)+\dfrac{2G^2JMb}{c^6r^4}(1+2\chi)-\dfrac{8G^2M \hbar}{5 \pi c^6 r^3}+\O(c^{-7}) \right] \equiv {\rm d}y f(y),
\end{equation}
and hence, once the integrals listed in Appendix \ref{Appendix_Integrals} have been exploited, the time propagation of the signal from the emitter position to the receiver becomes
\begin{equation}
\begin{split}
& \Delta t_{{\rm prop}} = \int_{-y_1}^{y_2} f(y) {\rm d}y  =  \dfrac{y_2+y_1}{c}+\dfrac{4 G M}{c^3} \log \left(\dfrac{y_2+y_1}{b}\right)+\dfrac{2GM}{c^3}\left(\dfrac{R}{b}\right)^2J_2 \pm  \dfrac{4GJ}{c^4b} \\
 & +\dfrac{3G^2M^2}{2c^5} \left(\dfrac{\pi}{b}-\dfrac{4}{(y_1+y_2)}+\dfrac{J_2 \,\pi R^2}{2b^3}\right) 
 \pm \dfrac{2 G^2 J Mb}{c^6} \left[\dfrac{\pi}{2b^3}+J_2\,R^2 \left(\dfrac{1}{2b^2(y_1y_2)^{3/2}}+\dfrac{3 \pi }{8 b^5} \right)   \right] - \dfrac{16G^2 M \hbar}{5 \pi c^6 b^2} + \O(c^{-7}).
\end{split}
\label{ShapiroTimeDelay_Quantum}
\end{equation}
The first line of Eq. (\ref{ShapiroTimeDelay_Quantum}) coincides with the result of Eq. (\ref{ShapiroTimeDelay_Kerr}). The second line contains new contributions: the first two factors are purely classical and represent corrections beyond the second post-Newtonian order, whereas the last term embodies the first-order quantum corrections. By considering a beam moving in the gravitational field produced by the Earth along the path linking the Lagrangian points $L_1$ and $L_2$, the quantum term occurring in (\ref{ShapiroTimeDelay_Quantum}) gives
\begin{equation}
\vert \Delta t_{{\rm prop}}^{({\rm q})} \vert  \simeq 9.7 \times 10^{-95} \; {\rm s}.
\label{Value_Quantum_Correction}
\end{equation}
Therefore, this correction is so tiny that it cannot be tested, in complete accordance with the ``observational spirit" of quantum gravity. However, as has been shown in Refs. \cite{B1,B1a,B2,B3}, in the low-energy regime the quantum theory of gravitation gives some effects regarding the position of the Earth-Moon Lagrangian points which could have the opportunity to be measured in the next future.

As we have pointed out before, in our model photons are made to follow a straight path and hence all contributions to time delay coming from the bending of light are completely neglected. However the difference $\Delta t$ between the lapse of coordinate time $t_{\rm s.t.}$  taken by a beam following a straight trajectory and a beam travelling on the real slightly curved route $t_{\rm c.t.}$ amounts to be \cite{MTW}
\begin{equation}
\dfrac{\Delta t} {t_{\rm s.t.}} =\dfrac{\vert t_{\rm s.t.}-t_{\rm c.t.}\vert}{t_{\rm s.t.}}\simeq ({\rm angle \; of \; deflection})^2.
\label{MTW_1}
\end{equation}
Therefore, as we said before, the phenomenon of path deflection gives contributions to time delay of order $\O(G^2M^2/c^6)$, making them comparable or even stronger than corrections appearing in (\ref{ShapiroTimeDelay_Quantum}). In other words, at the order displayed in Eq. (\ref{ShapiroTimeDelay_Quantum}) effects coming from the bending of light become quite considerable and hence they deserve to be mentioned.  

The deflection angle for a light beam evolving in Kerr geometry is given by \cite{Tartaglia2002}
\begin{equation}
\delta=\sqrt{\left(\dfrac{4GM}{c^2b}+\dfrac{4GJ_2\,MR^2}{c^2b^3}\right)^2+\left(\dfrac{4GJ}{c^3b^2}\right)^2},
\label{Deflection_angle}
\end{equation}
where the first term is the same as in Schwarzschild spacetime, the second one is due to the oblateness of the body of mass $M$, whereas the last one is related to angular momentum and hence it represents a gravitomagnetic effect. From Eqs. (\ref{MTW_1}) and (\ref{Deflection_angle}) we obtain (in the gravitational field of the Earth, $\Delta t_{{\rm prop}}^{\rm (s.t)}$ being given by Eq. (\ref{ShapiroTimeDelay_Quantum}))
\begin{equation}
\dfrac{\Delta t_{{\rm prop}}^{\rm (s.t)}-\Delta t_{{\rm prop}}^{\rm (c.t)}}{\Delta t_{{\rm prop}}^{\rm (s.t)}} \simeq \delta^2 \sim 10^{-18},
\label{Deflection_angle2}
\end{equation}
showing how contributions coming from light bending are far more significant than the first order quantum corrections (\ref{Value_Quantum_Correction}).  

It is interesting to note that quantum corrections do not change if we move to Schwarzschild geometry, i.e., the static case. Indeed, the quantum corrected Schwarzschild metric can be easily derived from Eq. (\ref{Quantum_Kerr}) by requiring that the spin-independent terms for a gravity-scalar and a gravity-fermion theory should be the same \cite{Bohr2003}. Therefore, setting aside the oblateness of the source mass, the hybrid scheme described before yields readily
\begin{equation}
\begin{split}
g_{00}^{({\rm S})}&=-1 + \dfrac{2GM}{c^2 r} (1+\chi)-\dfrac{2G^2 M^2}{c^4 r^2}(1+2\chi)-\dfrac{62 G^2 M \hbar}{15 \pi c^5 r^3} +\O(c^{-6}), \\
g_{ij}^{({\rm S})}&=\delta_{ij}\left[1+\dfrac{2GM}{c^2r}(1+\chi)+\dfrac{G^2M^2}{c^4r^2}(1+2\chi) + \dfrac{14 G^2 M \hbar}{15 \pi c^5 r^3}\right] 
 +\dfrac{r_i r_j}{r^2}\left( \dfrac{G^2M^2}{c^4r^2}+\dfrac{76G^2M\hbar}{15 \pi c^5 r^3}+\O(c^{-6})\right),\\
g_{0i}^{({\rm S})}& =0.
\end{split}
\label{Quantum_Schwarzschild}
\end{equation}
Since the off-diagonal terms, proportional to $\boldsymbol{J}$, enter the previous calculations only classically via Eq. (\ref{g_0y_squared}), we can easily conclude that the quantum corrected time delay within Schwarzschild geometry can be obtained from Eq. (\ref{ShapiroTimeDelay_Quantum}) by putting $J=0$, i.e., 
\begin{equation}
\begin{split}
\Delta t_{{\rm prop}}^{({\rm S})} & =  \dfrac{y_2+y_1}{c}+\dfrac{4 G M}{c^3} \log \left(\dfrac{y_2+y_1}{b}\right)\\
 +\dfrac{2GM}{c^3}\left(\dfrac{R}{b}\right)^2 J_2 +& \dfrac{3G^2M^2}{2c^5} \left(\dfrac{\pi}{b}-\dfrac{4}{(y_1+y_2)}+\dfrac{J_2 \,\pi R^2}{2b^3}\right) - \dfrac{16G^2 M \hbar}{5 \pi c^6 b^2} + \O(c^{-7}).
\end{split}
\label{ShapiroTimeDelay_QuantumSchwarzschild}
\end{equation}
The first line of (\ref{ShapiroTimeDelay_QuantumSchwarzschild}) is the same as Eq. (\ref{ShapiroTimeDelay_Schwarzschild}), where we did not consider oblateness. However, in the static case the calculation of the travel time can be simplified by invoking the relativistic version of Fermat principle, stating that all null curves between two points in {\it space} of extremal coordinate time interval $\Delta t$ represent null geodesics of spacetime \cite{MTW}. Despite that, it is obvious that all the arguments regarding deflected-path-effects leading to Eqs. (\ref{Deflection_angle}) and (\ref{Deflection_angle2}) are still valid also in the static framework. 

\section{Concluding remarks and open problems}

The development of a research programme in which we seek observable low-energy quantum gravity effects within solar system represents the origin of this paper. This dates back to Refs. \cite{B1,B1a,BattistaThesis,B2,B3} where some of us have proposed a model where quantum corrections to the position of the Earth-Moon Lagrangian points can be worked out. The possibility of measuring the predicted results represents a crucial step in our understanding of quantum phenomena affecting the gravitational field in the low-energy limit, since it could shed light on the eventual correctness of effective field theories approach. Indeed, such a pattern has been expansively employed in theoretical physics (as for example chiral perturbation theories witness), but so far an experimental confirmation has never been achieved. 

``LAGRANGE'' proposal fits sharply within such a scheme. Indeed, the configuration, rigidly rotating with the Earth, defined by the five Lagrangian points of the Sun-Earth system gives the opportunity to test fundamental physics effects at the scale of inner solar system. This property has already been exploited many times for
space missions, such as WMAP, the Herschel space observatory, Planck (all concluded) and now Gaia, in $L_2$; the Deep Space Climate Observatory, the Solar and Heliospheric Observatory (SOHO) and LISA Pathfinder, in $L_1$. The list is not exhaustive and many more missions are planned, directed again to $L_1$ or $L_2$. 

Therefore, in Sec. II we have performed a detailed calculation of the time delay experienced by an electromagnetic signal travelling along the straight path linking the Sun-Earth Lagrangian points $L_1$ and $L_2$ by exploiting the insights developed in Ref. \cite{Tartaglia2016}. In fact, this purely classic model represents the starting point towards the direct estimation of the Earth's quadrupole coefficient. This kind of measurement can be initiated by Earth, the delaying body, with all the advantages of an Earth-based Laboratory equipped with the best time-measuring apparatus to perform the experiment. In particular, optical clocks and lattice clocks based on $Sr$-atoms have reached outstanding fractional frequency instabilities down at a level of about $2 \times 10^{-16}/\sqrt{T}$ or less, $T$ being the integration time \cite{Sr-atoms}.

Section III is devoted to the evaluation of the quantum corrected version of the time delay presented in Sec. II. We exploit a pattern where quantum gravity is completely treated as an effective field theory. This model allows a natural separation of the (known) low-energy quantum effects from the (unknown) high-energy contributions. These leading corrections represent the first quantum modifications occurring in the gravitational field provided by a scheme where calculations are organised as an expansion in powers of the energy or inverse factors of the distance. Since they are independent of the eventual high-energy theory of gravity (they depend only on the massless degrees of freedom and their low energy couplings), these modifications are true predictions of quantum general relativity. Within this framework, gravity is a well behaved quantum field theory at ordinary energies and hence renormalizability issues can be overcome in such a way that (hopefully testable) quantum effects can be derived. Therefore, our first step towards the characterization of the quantum time delay consisted in the search of the most suitable spacetime geometry where our calculations could be performed. In fact, in Ref. \cite{Bohr2003} a quantum corrected version of (Schwarzschild and) Kerr metric is brilliantly derived through a rather involved Feynman-diagram-analysis. However, such a computation relies on quantum modifications occurring in the energy momentum tensor of a {\it point} particle, meanwhile our model takes into account both finite dimensions and deviations from the spheric shape of the source. Therefore, we have proposed a hybrid scheme where quantum fluctuations affect only the first term of the multipole expansion underlying the Newtonian potential (\ref{potential_U}), i.e., the monopole contribution $M$. This procedure ends up in the definition of the quantum corrected metrics (\ref{Quantum_Kerr}) and (\ref{Quantum_Schwarzschild}). They represent an original ``extended-source''-version of the original metrics of Ref. \cite{Bohr2003} (cf. Eq. (\ref{Bohr_Quantum_Kerr})) which, to the best of our knowledge, has never been studied before in the literature. In this way, we have proved that both for Kerr and Schwarzschild geometries the leading quantum modification to classical time delay occurring in the gravitational field of the Earth on the $L_1-L_2$ path reads as (see Eqs. (\ref{ShapiroTimeDelay_Quantum}), (\ref{Value_Quantum_Correction}), and (\ref{ShapiroTimeDelay_QuantumSchwarzschild}))
\begin{equation}
\Delta t_{{\rm prop}}^{({\rm q})}= -\dfrac{16G^2 M \hbar}{5 \pi c^6 b^2}   \simeq -9.7 \times 10^{-95} \; {\rm s}.
\label{Conclusion1}
\end{equation}
Unfortunately this deviation from the underlying classical theory has no chance to be tested, unlike, for example, the effects some of us predicted in Refs. \cite{B1,B1a,B2,B3}, where it is shown that Earth-Moon Lagrangian points are expected to undergo a displacement of the order of few millimetres from their classical position. Furthermore, we have proved that the result reported in Eq. (\ref{Conclusion1}) is completely hidden by those contributions to time delay generated by the bending of light, which have been neglected in our model. Nevertheless, we believe that our calculation can be of some interest because it shows how the path towards the possibility to test low-energy quantum gravity within solar system turns out to be highly demanding. Moreover, the framework described in this paper belongs to a wide class of quantum effects predicted in the literature which, despite having extremely tiny possibilities to be measured in the next future, reveals important insights into the as yet unkown quantum theory of gravitation. As an example, recently the authors of Ref. \cite{Pessina} have computed quantum gravitational corrections terms to the cosmic microwave background radiation anisotropy spectrum as they are found from a Born-Oppenheimer type of approximation from the Wheeler-DeWitt equation and they showed that these fluctuations are too small to be observed. 

In conclusion, the extreme smallness of the result (\ref{Conclusion1}) unveils that it remains an interesting open issue to look for some quantum phenomenon which can allow us to test low-energy quantum gravity within a quantization programme where the starting point is represented by effective-field-theory models of gravity.

\begin{appendix}

\section{Useful integrals} \label{Appendix_Integrals}

The computation of time propagation $\Delta t_{{\rm prop}}$ occurring in Eqs. (\ref{ShapiroTimeDelay_Schwarzschild}), (\ref{ShapiroTimeDelay_Kerr}), and (\ref{ShapiroTimeDelay_Quantum}) can be performed by exploiting the integrals listed below. All the expansions are carried out by assuming that $y_1\gg b$, $y_2 \gg b$, $y_1 \simeq y_2$ and are evaluated to first order in $b/y_1$ and $b/y_2$. We thus have ($r \equiv \sqrt{y^2+b^2}$):
\begin{equation}
\int_{-y_1}^{y_2} \dfrac{{\rm  d}y}{r}= \left. \log \left(y+ \sqrt{y^2+b^2} \right)\right\vert^{y_2}_{-y_1} \simeq 2 \log \left(\dfrac{y_2+y_1}{b} \right),
\end{equation}
\begin{equation}
\int_{-y_1}^{y_2} \dfrac{{\rm  d}y}{r^2}=\left. \dfrac{1}{b} \arctan\left(\dfrac{y}{b}\right)\right\vert^{y_2}_{-y_1} \simeq \dfrac{\pi}{b}-\dfrac{4}{(y_2+y_1)},
\label{int2}
\end{equation}
\begin{equation}
\int_{-y_1}^{y_2} \dfrac{{\rm  d}y}{r^3}=\left. \dfrac{y}{b^2 \sqrt{y^2+b^2}} \right\vert^{y_2}_{-y_1} \simeq \dfrac{2}{b^2},
\end{equation}
\begin{equation}
\int_{-y_1}^{y_2} \dfrac{{\rm  d}y}{r^4}=\left.\dfrac{1}{2b^3} \left[\dfrac{by}{(b^2+y^2)}+\arctan\left(\dfrac{y}{b}\right) \right]\ \right\vert^{y_2}_{-y_1} \simeq \dfrac{\pi}{2b^3},
\label{int4}
\end{equation}
\begin{equation}
\int_{-y_1}^{y_2} \dfrac{{\rm  d}y}{r^6}=\left. \left[\dfrac{y}{4b^2\left(b^2+y^2\right)^2}+\dfrac{3y}{8b^4\left(b^2+y^2\right)}+\dfrac{3\arctan\left(y/b\right)}{8b^5} \right] \right\vert^{y_2}_{-y_1} \simeq \dfrac{1}{2b^2 \left(y_1y_2\right)^{3/2}}+\dfrac{3 \pi}{8 b^5},
\label{int5}
\end{equation}
where the following relation has been exploited in Eqs. (\ref{int2}), (\ref{int4}), and (\ref{int5}):
\begin{equation}
\arctan x + \arctan \left( 1/x \right) = \pi/2, \;\;\;\;\; \;\;\;\;\; \forall\, x>0.
\end{equation}

\end{appendix} 
 
\acknowledgments
E. B., G. E. and L. D. F. are grateful to the Dipartimento di Fisica ``Ettore Pancini'' of Federico II University for
hospitality and support. The work of E. B., S. D., G. E. and A. G. has been supported by the 
INFN funding of the NEWREFLECTIONS experiment.


\begin{thebibliography}{99}

\bibitem{Shapiro64}
I. I. Shapiro, Fourth test of general relativity, Phys. Rev. Lett. {\bf 13}, 789 (1964).
\bibitem{Schwarzschild}
K. Schwarzschild, \"{U}ber das gravitationsfeld eines massenpunktes nach der Einsteinschen theorie (On the gravitational field of a point mass according to Einstein theory), Sitzungsberichte der K\"{o}niglich Preu{\ss}ischen Akademie der Wissenschaften (Berlin), Seite 189-196 (1916).
\bibitem{MTW}
C. W. Misner, K. S. Thorne, and J. A. Wheeler, {\it Gravitation} (W. H. Freeman and Company, San Francisco, 1973).
\bibitem{Tartaglia2002}
M. L. Ruggiero and A. Tartaglia, Gravitomagnetic effects, Nuovo Cim. B {\bf 117}, 743 (2002). \\
I. Ciufolini and F. Ricci, Time delay due to spin and gravitational lensing, Class. Quantum Grav. {\bf 19}, 3863 (2002).
\bibitem{Tartaglia2016}
A. Tartaglia, D. Lucchesi, M. L. Ruggiero, and P. Valko, Sun-Earth Lagrange reference for fundamental physics and navigation, arXiv: 1701.08217 (2017). 
\bibitem{Kerr}
R. P. Kerr, Gravitational field of a spinning mass as an example of algebraically special metrics, Phys Rev. Lett. {\bf 11}, 237238 (1963).
\bibitem{Ciufolini-Wheeler}
I. Ciufolini and J. A. Wheeler, {\it Gravitation and Inertia} (Princeton University Press, Princeton, 1995). 
\bibitem{LT}
J. Lense and H. Thirring, \"{U}ber den einfluss der eigenrotation der zentralk\"{o}rper auf die bewegung der planeten und monde nach der Einsteinschen gravitationstheorie (On the influence of the proper rotation of central bodies on the motions of planets and moons according to Einstein's theory of gravitation), Physikalische Zeitschrift. {\bf 19}, 156 (1918). \\
A. Einstein, {\it The meaning of relativity} (Princeton University Press, Princeton, 1923).
\bibitem{Pars}
L. A. Pars, {\it A Treatise on Analytical Dynamics} (Heinemann, London, 1965).
\bibitem{B1}
E. Battista and G. Esposito, Restricted three-body problem in effective-field-theory
models of gravity, Phys. Rev. D {\bf 89}, 084030 (2014).\\
\bibitem{B1a}
E. Battista and G. Esposito, Full three-body problem in effective-field-theory
models of gravity, Phys. Rev. D {\bf 90}, 084010 (2014); Phys. Rev. D {\bf 93}, 049901(E) (2016).
\bibitem{BattistaThesis}
E. Battista, Extreme regimes in quantum gravity, PhD thesis (Naples University, 2016); arXiv:1606.04259.
\bibitem{Donoghue}
J. F. Donoghue, General relativity as an effective field theory: the leading quantum corrections,
Phys. Rev. D {\bf 50}, 3874 (1994). \\
A. A. Akhundov, S. Bellucci, and A. Shiekh, Gravitational interaction to one loop in effective
quantum gravity, Phys. Lett. B {\bf 395}, 16 (1997).\\
J. F. Donoghue, The effective field theory treatment of quantum gravity, AIP Conf. Proc.
{\bf 1483}, 73 (2012).\\
N. E. J. Bjerrum-Bohr, J. F. Donoghue, B. R. Holstein, L. Plant\'e, P. Vanhove, Light-like scattering in quantum gravity, J. High Energ. Phys. {\bf 11} (2016) 117.\\
J. F. Donoghue, M. M. Ivanov, A. Shkerin, EPFL Lectures on General Relativity as a Quantum Field Theory, arXiv: 1702.00319 [hep-th]. 
\bibitem{B2}
E. Battista, S. Dell'Agnello, G. Esposito and J. Simo, Quantum effects on Lagrangian points and
displaced periodic orbits in the Earth-Moon system, Phys. Rev. D {\bf 91}, 084041 (2015);
Phys. Rev. D {\bf 93}, 049902(E) (2016).
\bibitem{B3}
E. Battista, S. Dell'Agnello, G. Esposito, L. Di Fiore, J. Simo and A. Grado, Earth-Moon Lagrangian 
points as a test bed for general relativity and effective field theories of gravity,
Phys. Rev. D {\bf 92}, 064045 (2015); Phys. Rev. D {\bf 93}, 109904(E) (2016).
\bibitem{Scherer}
S. Scherer, Introduction to Chiral Perturbation Theory, Adv. Nucl. Phys. {\bf 27}, 277 (2003). 
\bibitem{'tHooft1974}
G. 't Hooft and M. Veltman, One-loop divergencies in the theory of gravitation, Ann. Ist. H. Poincar\'e A {\bf 20}, 69 (1974).
\bibitem{Bunch}
T. S. Bunch and L. Parker, Feynman propagator in curved spacetime: A momentum-space representation, Phys. Rev. D {\bf 20}, 2499 (1979). 
\bibitem{DW}
B. S. DeWitt, {\it The Global Approach to Quantum Field Theory} (Clarendon Press, Oxford, 2003).
\bibitem{D94}
J. F. Donoghue, Leading quantum correction to the Newtonian potential, Phys. Rev. Lett.
{\bf 72}, 2996 (1994).
\bibitem{D2003}
N. E. J. Bjerrum-Bohr, J. F. Donoghue, and B. R. Holstein, Quantum gravitational corrections
to the nonrelativistic scattering potential of two masses, Phys. Rev. D {\bf 67}, 084033
(2003).
\bibitem{Bohr2003}
N. E. J. Bjerrum-Bohr, J. F. Donoghue, and B. R. Holstein, Quantum corrections to the Schwarzschild and Kerr metrics, Phys. Rev. D {\bf 68}, 084005 (2003).
\bibitem{D2002}
J. F. Donoghue, B. R. Holstein, B. Garbrecht, and T. Konstandin, Quantum corrections to the Reissner-Nordstr\"{o}m and Kerr-Newman metrics, Phys. Lett. B {\bf 529}, 132 (2002).
\bibitem{Boyer-Lindquist}
R. H. Boyer and R. W. Lindquist, Maximal Analytic Extension of the Kerr Metric, J. Math. Phys. {\bf 8}, 265 (1967).\\
L. D. Landau and E. M. Lifshitz, {\it The Classical Theory of Fields} (Fourth revised English edition) (New York, Pergamon Press, 1975).
\bibitem{Reigber}
C. Reigber, R. Schimdt, F. Flechtner {\it et al.}, An Earth gravity field model complete to degree and order 150 from GRACE: EIGEN- GRACE02S., J. Geodyn. {\bf 39}, 110 (2005). \\
M. Cheng, B. D. Tapley, and J. C. Ries, Deceleration in the Earths oblateness, J. Geophys. Res. {\bf 118}, 740747 (2013). 
\bibitem{Sr-atoms}
A. Al-Masoudi {\it et al.}, Noise and instability of an optical lattice clock, Phys. Rev. A {\bf 92}, 063814, (2015).\\
T. L. Nicholson {\it et al.}, Systematic evaluation of an atomic clock at $2 \times 10^{-18}$ total uncertainty, Nat. Comm. {\bf 6}, 6896 (2015).
\bibitem{Pessina}
D. Bini, G. Esposito, C. Kiefer, M. Kr\"{a}mer, and F. Pessina, On the modification of the cosmic microwave background anisotropy spectrum from canonical quantum gravity, Phys. Rev. D {\bf 87}, 104008 (2013). 
\end{thebibliography}
\end{document}